\documentstyle[aps,prb,epsbox]{revtex}
\draft
\begin{document}

% for two column
%\twocolumn[
%\hsize\textwidth\columnwidth\hsize\csname@twocolumnfalse\endcsname

\title{
Pump Built-in Hamiltonian Method for Pump-Probe Spectroscopy
}

\author{Jun-ichi Inoue 
and Akira Shimizu}
\address{
Department of Basic Science, University of Tokyo, 
3-8-1 Komaba, Meguro-ku, Tokyo 153-8902
\\
Core Research for Evolutional Science and Technology
 (CREST), JST
}
\date{\today}
\maketitle

\def\lesssim{\ ^< \!\!\!\!_{\sim}\ }
\def\gtrsim{\ ^> \!\!\!\!_{\sim}\ }

\begin{abstract}
We propose a new method of calculating
nonlinear optical responses of interacting electronic systems.
In this method,
the total Hamiltonian (system
+ system-pump interaction)
is transformed into a different form that (apparently)
does not have a system-pump interaction.
The transformed Hamiltonian, which we call
the pump built-in Hamiltonian,
has parameters that depend on the strength of the
pump beam.
Using the pump built-in Hamiltonian, we can
calculate nonlinear
responses (responses to probe beams as a function of the pump beam) by
applying the {\em linear} response theory.
We demonstrate the basic idea of this new method
by applying it to a one-dimensional, two-band model, in the case
the pump excitation is virtual (coherent excitation).
We find that the exponent of the Fermi edge singularity
varies with the pump intensity.
\end{abstract}
%]
% for two column

%\section{Introduction}
Nonlinear optics for semiconductors is one of active
fields in solid state physics \cite{Shen}.  
Recently, research interests of nonlinear optics are
extended to Mott-Hubbard and charge-transfer insulators \cite{Imada},
for which electron-electron interactions play central roles.  
The nonlinear optical responses, especially pump-probe spectroscopy,
from these systems are experimentally studied, and quite interesting
phenomena are reported \cite{Ogasawara}.  
However, it is practically impossible to apply the standard method of
calculation of $\chi^{(3)}$ to such systems, because it is
quite difficult to obtain the wavefunctions of the ground states and the excited
state.  

In this letter, we propose a new approach, in which the total Hamiltonian $H(t)$ (system 
Hamiltonian $H_{sys}$ + system-pump interaction $H_{sys-pump}(t)$)
is transformed into a different form that (apparently) does not have a system-pump
interaction.
The transformed Hamiltonian is called a ``pump built-in Hamiltonian''.  
If the pump built-in Hamiltonian is tractable, one can calculate nonlinear
responses (responses to probe beams as a function of the pump beam) by applying
the {\em linear} response theory \cite{Kubo} 
to the pump built-in Hamiltonian.\cite{similar}  
We here present the main idea by explicitly constructing the pump built-in
Hamiltonian in a two-band model when the pump excitation is virtual
(in the sense we explain below).
It is shown that we can include effects of
the pump beam as modification of the parameters of 
the original Hamiltonian in some cases.  
Namely, the system with a pump beam 
is equivalent to the system
without the pump beam with different values of parameters.
Using the pump built-in Hamiltonian, we successfully evaluate 
the Fermi edge singularity of the absorption of the probe beam. 
It is shown that 
the exponent of the singularity varies 
as a function of the pump intensity.
% ; this effect may be called
% the nonlinear Fermi edge singularity.
This success indicates that 
the pump built-in Hamiltonian method 
could have potential to analyze nonlinear optical responses
of various systems in which electron-electron interactions
play important roles.
% from, {\em e.g.}, Mott-Hubbard insulators.  

We assume that the pump excitation is ``virtual'', i.e., 
the quantum-mechanical coherence is preserved when the pump beam 
is shining the sample. 
% That is, the virtual excitation here means coherent excitation, 
% in contrast to the ``real excitation'' which means incoherent excitation.
The virtual excitation is realized when the spectrum of the pump beam 
does not overlap with the absorption spectrum of the system.
If the state vector $|\Psi(t) \rangle$ of the system is 
initially the ground state $|G \rangle$ 
of $H_{sys}$ before the pump beam is on, 
it evolves adiabatically the ``instantaneous ground state''
of 
$
H(t) = H_{sys}+H_{sys-pump}(t)
$ in the rotating frame.
Namely,  
\begin{equation}
|\Psi(t) \rangle
=
|G[{\cal E}(t)] \rangle,
\end{equation}
where $|G[{\cal E}(t)] \rangle$ denotes 
% the ``instantaneous ground state''  of 
% $
% H(t)% = H_{sys}+H_{sys-pump}(t)
% $, 
% i.e., 
the ground state of $H(t)$ in the rotating frame when $t$ is regarded as
a parameter.  
It is a function of the instantaneous value 
of the envelope function ${\cal E}(t)$ of the pump beam at time $t$.
% Hence, the pump beam just deforms the wavefunction of the system, 
% and $|\Psi \rangle = |$
Therefore, 
it is sufficient to calculate the {\em linear} response 
of $|G[{\cal E}] \rangle$ 
to a probe beam 
{\em for each fixed value} of ${\cal E}$; 
it gives the {\em nonlinear} response of the system at a time $t$ 
that satisfies ${\cal E} = {\cal E}(t)$.
We denote $H(t)$ for ${\cal E}={\cal E}(t)$ simply as $H$.  

% Besides, another theoretical benefit to consider virtual excitation is that it is not needed to
% discuss relaxation processes explicitly.  

%--------------------------------------------------------------------------
%\section{body}

To demonstrate the basic idea of the 
pump built-in
Hamiltonian method, 
we here consider a simple model;
% In order to demonstrate that we can obtain a pump built-in
% Hamiltonian, we consider 
a one-dimensional, two-band   
model which has % the Fermi wavenumber $k_{F}$ and 
a finite band gap $E_{g}$.  
The one-dimensional system may be realized, e.g., in a quantum wire
made of semiconductors.
The two bands, which we call ``electron'' and ``hole'' bands, 
can be either the lowest conduction subband and the highest 
valence subband, or
the second and the first subbands of the conduction band.
The latter case may be more suitable to realize
the virtual excitation\cite{SY}.
We assume that the pump beam is polarized in the direction 
normal to the quantum wire direction, whereas
the probe beam is polarized parallel to the wire.
Let $c_{k,\sigma}\,(c^{\dagger}_{k,\sigma})$ and
$d_{-k-\sigma}\,(d^{\dagger}_{-k-\sigma})$ be the Fermi annihilation (creation)
operators for an electron and a hole, respectively.  
We take the model Hamiltonian used in Refs.\,\cite{Aihara} and \cite{Comte}, 
which has the following form in the rotating frame \cite{Galitzkii};
\begin{eqnarray}
H&=&H_{1}+H_{2},
\nonumber\label{H} \\
H_{1}&=&\sum_{k,\sigma}\left(
E_{c}(k)c^{\dagger}_{k\sigma}c_{k\sigma}+
E_{d}(k)d^{\dagger}_{-k-\sigma}d_{-k-\sigma}\right)
\\
&&+\sum_{k,\sigma}\lambda(k)
(c_{k\sigma}d_{-k-\sigma}+d^{\dagger}_{-k-\sigma}c^{\dagger}_{k\sigma}),
\label{H_{1}}
\\
H_{2}&=&\frac{1}{2}\sum_{\sigma,\sigma'}\sum_{k,k',q}
U(q)\Bigl(
c^{\dagger}_{k+q\,\sigma}c^{\dagger}_{k'-q\,\sigma'}c_{k'\,\sigma'}c_{k\,\sigma}
\nonumber \\
&&+
d^{\dagger}_{-k-q\,-\sigma}d^{\dagger}_{-k'+q\,-\sigma'}d_{-k'\,-\sigma'}d_{-k\,-\sigma}
\nonumber \\
&&-2
c^{\dagger}_{k+q\,\sigma}c_{k\,\sigma}d^{\dagger}_{-k'+q\,-\sigma'}d_{-k'\,-\sigma'}\Bigr).
\label{H_{2}}
\end{eqnarray}
Here, 
the electron and hole energies in the effective-mass approximation are
taken as
\begin{eqnarray}
 E_{c}(k)&=&\frac{k^{2}}{2m_{c}}+E_{g}-\omega_{p},\\
 E_{d}(k)&=&\frac{k^{2}}{2m_{d}},
\end{eqnarray}
respectively, where $m_{c}$ and $m_{d}$ are the electron ({\em c}) and hole ({\em d}) effective
masses, and $\omega_p$ is the central frequency of the pump beam.
The $\lambda(k)$ is the product of ${\cal E}$ and the
interband transition dipole.  
The $U(q)>0$ represents the interaction strength between particles.  
The detuning energy $\Delta\equiv E_{g}-\omega_{p}$ 
is assumed positive and large enough to satisfy the condition 
of the virtual excitation.
To simplify the calculations, we assume that 
the initial state $|G\rangle$ is two Fermi seas of 
electrons and holes with the same Fermi wavenumber $k_F$.
This may be realized, e.g., by
creating electrons and holes by another optical beam;
the pump beam is shined after the carriers cool down. 
To further simplify the calculations, we consider the case where
\begin{equation}
E_g \gg |E_{i}(k_{F})-E_{i}(0)| \gg |U|.
\label{condition}
\end{equation}
Under these conditions,  
spontaneous band mixing, 
such as the instability toward an excitonic insulator, 
are unfavorable.
Moreover, 
many-body scatterings occur only among electrons and holes near 
their Fermi points.
Furthermore,  
we can neglect interband scatterings (hence such terms have not been included in $H$).

Our purpose is to construct Hamiltonian in which the pump term, the
second term in Eq.\,(\ref{H_{1}}), is eliminated.  
To this end, we first diagonalize $H_{1}$ by the Bogoliubov transformation
\begin{eqnarray}
\tilde{c}_{k\sigma}
&=&\cos\theta_{k}c_{k\sigma}+\sin\theta_{k}d^{\dagger}_{-k-\sigma},
\label{bogc}\\
\tilde{d}_{-k\-\sigma}
&=&\cos\theta_{k}d_{-k-\sigma}-\sin\theta_{k}c^{\dagger}_{k\sigma},
\label{bogd}
\end{eqnarray}
as
\begin{equation}
H_{1}=\sum_{k,\sigma}\tilde{E}_{+}(k)
\tilde{c}^{\dagger}_{k\sigma}\tilde{c}_{k\sigma}
+\tilde{E}_{-}(k)\tilde{d}^{\dagger}_{-k-\sigma}\tilde{d}_{-k-\sigma}
\equiv \tilde{H}^{0}_{1},
\end{equation}
where
\begin{eqnarray}
 \tilde{E}_{\pm}(k)&=&\sqrt{E_{+}(k)^{2}+\lambda(k)^{2}}\pm E_{-}(k),
\\
\cos2\theta_{k}&=&E_{+}(k)/\sqrt{E_{+}(k)^{2}+\lambda(k)^{2}},
\\
\sin2\theta_{k}&=&-\lambda(k)/\sqrt{E_{+}(k)^{2}+\lambda(k)^{2}},
\\
E_{\pm}(k)&\equiv&[E_{c}(k)\pm E_{d}(k)]/2.
\end{eqnarray}
Note that $\tilde{E}_{+}-E_{c}$ and $\tilde{E}_{-}-E_{d}$ represent the
optical Stark shifts of the (non-interacting) electron and hole,
respectively.  
We then rewrite $H_{2}$
using the transformed operators, $\tilde{c}$, $\tilde{c}^{\dagger}$, 
$\tilde{d}$, and $\tilde{d}^{\dagger}$.  
Substituting these operators in $H_{2}$, and ordering them
in their normal order, we obtain
$H_{2}=\tilde{H}_{1}'+\tilde{H}_{2}+\tilde{H}_{2}'$.  
Here $\tilde{H}_{1}'$ is the one-body part, whereas $\tilde{H}_{2}$ and
$\tilde{H}'_{2}$ are many-body interactions.  
The $\tilde{H}_{2}$ consists of terms of the forms of
$\tilde{c}^{\dagger}\tilde{c}^{\dagger}\tilde{c}\tilde{c}$,
$\tilde{d}^{\dagger}\tilde{d}^{\dagger}\tilde{d}\tilde{d}$, and
$\tilde{c}^{\dagger}\tilde{d}^{\dagger}\tilde{d}\tilde{c}$.  
On the other hand, $\tilde{H}'_{2}$ consists of other combinations of
$\tilde{c}$, $\tilde{c}^{\dagger}$, $\tilde{d}$, and
$\tilde{d}^{\dagger}$.  
Needless to say, only gauge invariant combinations appear.  
Collecting all the one-body terms, we obtain the one-body part of the
transformed Hamiltonian, $\tilde{H}_{1}\equiv\tilde{H}_{1}^{0}+\tilde{H}_{1}'$, in the following form:  
\begin{eqnarray}
\tilde{H}_{1}&=&\sum_{k,\sigma}\left(
\tilde{E}_{c}(k)\tilde{c}^{\dagger}_{k\sigma}\tilde{c}_{k\sigma}+
\tilde{E}_{d}(k)\tilde{d}^{\dagger}_{-k-\sigma}\tilde{d}_{-k-\sigma}\right)
\nonumber \\
&&+\sum_{k,\sigma}\tilde{\lambda}(k)(\tilde{c}_{k\sigma}\tilde{d}_{-k-\sigma}
                 +\tilde{d}^{\dagger}_{-k-\sigma}\tilde{c}^{\dagger}_{k\sigma}).  
\end{eqnarray}
On the other hand, the form of $\tilde{H}_{2}$ is
\begin{eqnarray}
\tilde{H}_{2}&=&\frac{1}{2}\sum_{\sigma,\sigma'}\sum_{k,k',q}
\tilde{U}(q,k,k')\Bigl(
\tilde{c}^{\dagger}_{k+q\,\sigma}\tilde{c}^{\dagger}_{k'-q\,\sigma'}
\tilde{c}_{k'\,\sigma'}\tilde{c}_{k\,\sigma}
\nonumber \\
&&+
\tilde{d}^{\dagger}_{-k-q\,-\sigma}\tilde{d}^{\dagger}_{-k'+q\,-\sigma'}
\tilde{d}_{-k'\,-\sigma'}\tilde{d}_{-k\,-\sigma}
\nonumber \\
&&-2
\tilde{c}^{\dagger}_{k+q\,\sigma}\tilde{c}_{k\,\sigma}
\tilde{d}^{\dagger}_{-k'+q\,-\sigma'}\tilde{d}_{-k'\,-\sigma'}\Bigr).
\label{H2f}
\end{eqnarray}
We consider two simplified
cases:  
(a) $U(0)\gg U(2k_{F})$ \cite{Ogawa} and (b) $U(0)\simeq
U(2k_{F})$.  

Firstly, we consider case (a), for which backward scattering is
negligible.  
In this case, under the condition (\ref{condition}), we can take $k=\pm
k_{F}$, $k'=\pm k_{F}$, and $q\simeq 0$.  
Therefore $\tilde{U}(q, k, k')$ is represented by a single parameter $\tilde{U}(0,\pm
k_{F},\pm k_{F})\equiv\tilde{U}$. 
We then find that $\tilde{H}'_{2}$ is negligible.  
Comparing the explicit forms of $H_{2}$ and $\tilde{H}_{2}$, we also find
that $\tilde{H}_{2}$ has exactly the same form as $H_{2}$, and that the
parameters are changed as
%see 1999/02/02p2
\begin{eqnarray}
&&\tilde{E}_{c/d}(k)
=\tilde{E}_{+/-}(k)  +\frac{U}{2}\sin^{2}\theta_{k}
\nonumber \\
&&\qquad-\frac{3U}{2}\sin^{2}\theta_{k}\cos 2\theta_{k}
 +\frac{U}{2}\sin^{2} 2\theta_{k},
\label{reEc}
\\
&&\tilde{\lambda}(k)
=-\frac{U}{2}\sin 2\theta_{k},
\label{rel}
\end{eqnarray}
and $\tilde{U}=U$.
\label{reU}
Furthermore, the Fermi wavenumber of the transformed
system, $\tilde{k}_{F}$, is equal to $k_{F}$ because we have assumed that
$|E(k_{F})-E(0)|\gg |U|$ and that the excitation is virtual: 
in the virtual excitation, the densities of ``dressed particles''
$(\tilde{c}$ and $\tilde{d})$ are conserved.

Although our purpose is to eliminate the pump term,
a new pump term has been generated through the many-body interactions, 
with the coefficient $\tilde{\lambda}(k)$.
To eliminate this term, 
we perform Bogoliubov transformations successively: 
since the form of $\tilde{H}_{2}$ is the same as $H_{2}$, 
we can perform the same procedure successively.  
The mapping functions from $\{E_{c}, E_{d}, \lambda\} \to \{\tilde E_{c}, \tilde
E_{d}, \tilde\lambda\}$, Eqs.\,(\ref{reEc}) and (\ref{rel}), are
denoted by $f_{j}(E_{c}, E_{d}, \lambda)\ (j=1,2,3)$.
Then the recursion relations between steps $n$ and $n+1$ are written as follows:
\begin{eqnarray}
 E^{(n+1)}_{c}&=&f_{1}(E^{(n)}_{c}, E^{(n)}_{d}, \lambda^{(n)}),
\label{iteration1}
\\
 E^{(n+1)}_{d}&=&f_{2}(E^{(n)}_{c}, E^{(n)}_{d}, \lambda^{(n)}),
\label{iteration2}
\\
 \lambda^{(n+1)}&=&f_{3}(E^{(n)}_{c}, E^{(n)}_{d}, \lambda^{(n)}).
\label{iteration3}
\end{eqnarray}
These recursion relations are solved numerically.  
\begin{figure}[t]
\begin{center}
\epsfile{file=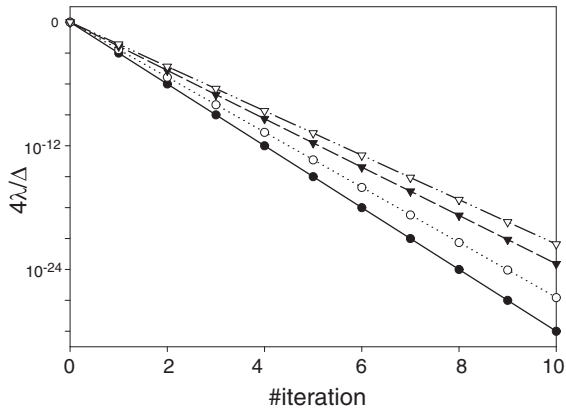,scale=0.5}
\end{center}
\caption{
The semi-log plot of $\lambda^{(n)}$ versus the number of
 iterations in case (a) $U(0)\gg U(2k_{F})$, for
 several values of $4U/\Delta=$6, 4, 2, and 1 (from the top to the bottom).  
The $\lambda$ of the original system is fixed.  
}
\end{figure}
Figure\,1 shows the semi-log plot of $\lambda^{(n)}$ for several
values of $U/\Delta$, where the pump-system coupling strength of the original
system, $\lambda$, is fixed.  
It is found that $\lambda^{(n)}$ decreases
exponentially with increasing the number of iterations, converging to zero for all
values of $U/\Delta$.  
If we define the decay exponent $\mu$ by
$\lambda^{(n)}\propto
\exp[-\mu n]$,
the plot shows that the larger the forward scattering strength is, the
smaller the decay exponent is.  
\begin{figure}[t]
\begin{center}
 \epsfile{file=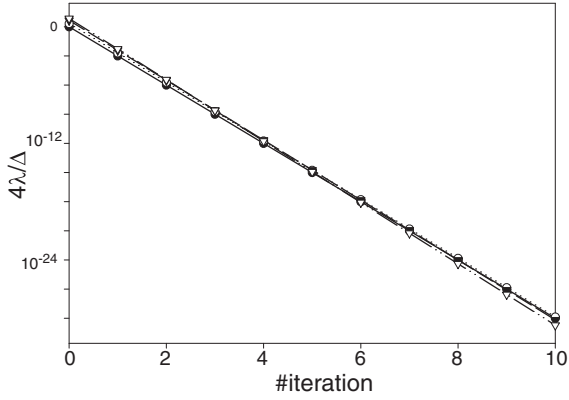,scale=0.5}
\end{center}
\caption{
The semi-log plot of $\lambda^{(n)}$ versus the number of
 iterations  in case (a) $U(0)\gg U(2k_{F})$, for
 several values of $4\lambda/\Delta$=6, 4, 2, and 1 (from the top
 to the bottom on the left vertical axis).  
The value $4U/\Delta=1$ is fixed.  
}
\end{figure}
Figure\,2 shows the semi-log plot of $\lambda^{(n)}$ for several
values of $\lambda/\Delta$, where $4U/\Delta=1$ is fixed.  
It shows that $\lambda^{(n)}$ converges to zero in the limit of
$n\to\infty$ for all values of $\lambda/\Delta$.  
The plot also shows that the decay exponent is almost independent of
$\lambda$ of the original system.  
We thus find that $\lambda^{(n)}$ converges to zero after
iterations for any values of $U>0$ and $\Delta>0$,
and that the decay exponent depends only on the strength of
the forward scattering.

Next the recursion relations for the kinetic energies are considered.  
Figure\,3 shows that the kinetic energies of the electron and the hole in the
limit of $n\to \infty$, for $4U/\Delta=1$ and $4\lambda/\Delta=1$.  
The curvatures of the dispersions become slightly smaller than the
original ones (dotted lines).  
Qualitatively similar results are obtained for other values of
$U/\Delta>0$.  
We thus conclude that the elimination of the pump
term makes the masses of electron and hole heavier.  

Therefore the pump term is eliminated in the limit of $n\to \infty$, and  
we obtain the pump built-in Hamiltonian in case (a).

\begin{figure}[t]
\begin{center}
\epsfile{file=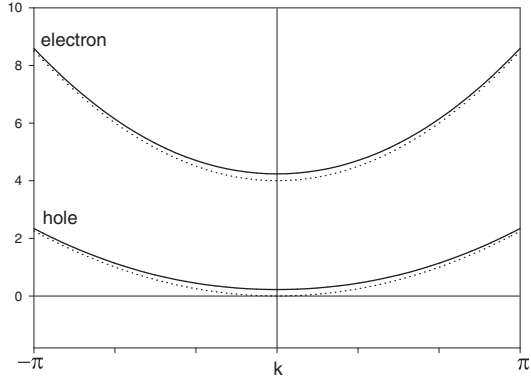,scale=0.5}
\end{center}
\caption{
The dispersion curves of the electron and the hole (black lines) in the
 limit of $n\to\infty$ in case (a) $U(0)\gg U(2k_{F})$.  
The dotted lines show the initial dispersions.  
}
\end{figure}
%-------------------------------------------------------------------------
Secondly, we consider case (b), where $U(0)\simeq U(2k_{F})$.  
This situation may be realized, e.g., for a screened interaction if
the screening length is larger than $1/k_{F}$.  
By dropping the wave number dependence
of the interaction strength, we find
\begin{eqnarray}
\tilde{H}_{2}&=&\frac{U}{2}\sum_{\sigma,\sigma'}\sum_{k,k',q}
\Bigl(
\tilde{c}^{\dagger}_{k+q\,\sigma}\tilde{c}^{\dagger}_{k'-q\,\sigma'}
\tilde{c}_{k'\,\sigma'}\tilde{c}_{k\,\sigma}
\nonumber \\
&&+
\tilde{d}^{\dagger}_{-k-q\,-\sigma}\tilde{d}^{\dagger}_{-k'+q\,-\sigma'}
\tilde{d}_{-k'\,-\sigma'}\tilde{d}_{-k\,-\sigma}
\nonumber \\
&&-2
\tilde{c}^{\dagger}_{k+q\,\sigma}\tilde{c}_{k\,\sigma}
\tilde{d}^{\dagger}_{-k'+q\,-\sigma'}\tilde{d}_{-k'\,-\sigma'}\Bigr),
\label{H2h}
\end{eqnarray}
and $\tilde{H}'_{2}=0$.  
As in case (a), we obtain
\begin{eqnarray}
&& \tilde{E}_{c/d}(k)
=\tilde{E}_{+/-}(k)+\frac{U}{2}I_{2}
\nonumber \\
&&\qquad  -\frac{3U}{2}I_{2}\cos 2\theta_{k}
+U I_{1}\sin 2\theta_{k},
\\
&&\tilde{\lambda}(k)
=-U I_{2}\sin 2\theta_{k}
                     -U I_{1}\cos 2\theta_{k},
\end{eqnarray}
and $\tilde{U}=U$.
The $I_{1(2)}$ are defined by
$
 I_{1}\equiv(1/2)\sum_{k}\sin 2\theta_{k}
$ and
$
 I_{2}\equiv\sum_{k}\sin^{2}\theta_{k}
$.  
From successive transformations, the recursion relations corresponding to
Eqs.\.(\ref{iteration1})--(\ref{iteration3}) are obtained.  
Figure 4 is the semi-log plot of $\lambda^{(n)}$ as a function of the
number of iterations.  
One can see that $\lambda$ converges to zero exponentially as in case
(a), although the decay exponent is smaller than that of case (a). 
The plot corresponding to Fig.\,3 is also obtained, which yields the same
conclusion that the elimination of the pump beam makes the
effective masses of electron and hole slightly heavier.  
We thus obtain the pump built-in Hamiltonian in the limit of
$n\to\infty$ also in case (b).  
\begin{figure}[t]
\begin{center}
\epsfile{file=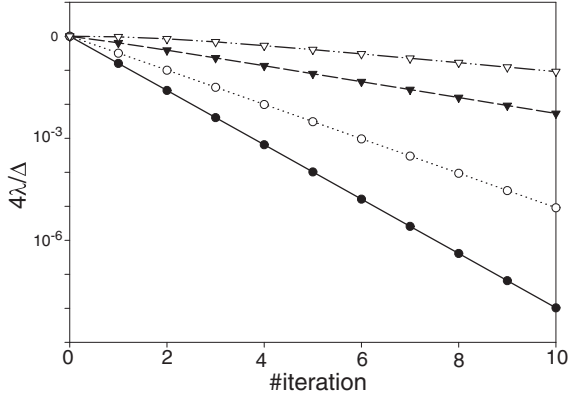,scale=0.5}
\end{center}
\caption{
The semi-log plot of $\lambda^{(n)}$ against the number of
 iterations in case (b) $U(0)\simeq U(2k_{F})$, for
 several values of $4U/\Delta=$6, 4, 2, and 1 (from the top to the bottom).  
The $\lambda$ of the original system is fixed.  
}
\end{figure}

\medskip
Since we have constructed the pump built-in Hamiltonian for both cases
(a) and (b), we can discuss
{\em nonlinear} optical responses by applying the 
{\em linear} response theory, 
the Kubo formula. 
For example, the absorption of the probe beam in the presence of a pump
beam is given by the real part of \cite{Mahan}
\begin{equation}
 \sigma({\bf q}, \omega)=\frac{1}{\omega}\int^{t}_{-\infty}dt' e^{i\omega(t-t')}
\langle [j^{\dagger}({\bf q},t), j({\bf q}, t')]\rangle,
\end{equation}
where $j({\bf q},t)$ is the current operator, 
and $\langle\cdots\rangle$ denotes the expectation value in 
the ground state $|G[{\cal E}] \rangle$ of the pump built-in Hamiltonian.  
For case (a), following Refs.\,(\cite{Ogawa}) and
(\cite{Otani}), we evaluate the power-law singularity of the
absorption of the probe beam;
$I(\omega)\sim|\omega-E_{F}|^{\eta({\cal E})-1}$.  
It is found that the exponent $\eta({\cal E})$ varies 
as a function of the intensity 
of the pump beam.  
The explicit form of $\eta({\cal E})$ is obtained by 
applying the bosonization technique, 
by which pump built-in Hamiltonian is diagonalized in terms of the charge $(j=1)$ and
spin $(j=2)$ excitations. 
The result is
\begin{equation}
 \eta({\cal E})=
\frac{\pi}{2}\sum_{i=\atop\{\tilde{e},\tilde{d}\}}\sum_{j=\atop\{1,2\}}
\left[
\frac{(u^{i}_{j})^{2}}{\sqrt{g^{i}_{j}}}
+(v^{i}_{j})^{2}\sqrt{g^{i}_{j}}
-2u^{i}_{j}v^{i}_{j}
\right],
\end{equation}
where
$[u^{\tilde{c}}_{j},u^{\tilde{d}}_{j}]^{t}
={\cal R}(\varphi_{j})[(v^{c*}_{F})^{1/2},(v^{d*}_{F})^{1/2}]^{t}$,
$[v^{\tilde{c}}_{j},v^{\tilde{d}}_{j}]^{t}
={\cal R}(\varphi_{j})[(v^{c*}_{F})^{-1/2},(v^{d*}_{F})^{-1/2}]^{t}$,
\begin{eqnarray}
 g^{\tilde{c}/\tilde{d}}_{j}&=&\Bigl[(1/2)
v^{c*}_{F}g^{c}_{j}+v^{d*}_{F}g^{d}_{j}
.\nonumber \\
&\pm&\sqrt{
(v^{c*}_{F}g^{c}_{j}-v^{d*}_{F}g^{d}_{j})^{2}+4\xi_{j}v^{c*}_{F}v^{d*}_{F}
}
\Bigr]
,
\end{eqnarray}
$g^{c(d)}_{1}=v^{c(d)*}_{F}+ U/\pi$, $g^{c(d)}_{2}=v^{c(d)*}_{F}-U/\pi$, 
$\xi_{1}=-2U/\pi$, and $\xi_{2}=0$.  
The ${\cal R}(\varphi_{j})$ is the $2\times2$ rotation matrix of angle
$\varphi_{j}$, where
\begin{equation}
\varphi_{c}
=(1/2)\tan^{-1}[2\xi_{1}\sqrt{v^{c*}_{F}v^{d*}_{F}}/(v^{c*}_{F}g^{c}_{1}-v^{d*}_{F}g^{d}_{1})],
\end{equation}
and
$\varphi_{2}=0$.
The $\eta({\cal E})$ is determined by the Fermi velocity
and the interaction strength in the pump built-in Hamiltonian.  
In the model under consideration, 
the Fermi velocity in the pump built-in
Hamiltonian, $v^{*}_{F}$, is smaller than the one without the
pump beam, $v_{F}$,  whereas the interaction strength is unchanged.
Hence, 
$\eta({\cal E})>\eta$, which should be observable in pump-probe
experiments on quantum wires.  
Since $|E(k_{F})-E(0)|\gg |U|$ is assumed, the difference between
$v^{*}_{F}$ and $v_{F}$ is small, hence 
the change of the exponent is also small.
If we take other models, however, 
we expect a larger change.

Finally, we mention another significance of the pump built-in Hamiltonian method.
Sato and coworkers recently observed a phase transition induced by 
a pump beam\cite{Sato}.
Miyashita and coworkers recently suggested that 
this phenomenon can be explained if one {\em assumes} that 
parameters in the Hamiltonian varies as a function of 
the pump intensity\cite{Miyashita}.
Since the pump built-in Hamiltonian has this property, 
the pump built-in Hamiltonian method may provide 
for the microscopic foundation of their assumption.
However, 
we do not exclude other possibilities, such as a thermal effect, 
for the origin of 
the changes of the parameters in the experiment of Ref.\ \cite{Sato}.  
Further studies are needed for settling it.


\begin{thebibliography}{99}
\bibitem{similar}
Similar methods were used, e.g., in refs.\ \cite{previous}.
However, these theories correspond to $n=1$ (no iteration) 
of the present work. Hence, actually, the pump term was not eliminated
as we have shown in eq.\ (\ref{rel})
\end{thebibliography}
\end{document}